\documentclass[aps,prl,twocolumn,superscriptaddress,showpacs,floatfix]{revtex4}

\usepackage{graphicx}
\usepackage{dcolumn}
\usepackage{bm}
\usepackage{stmaryrd}
\usepackage{latexsym}
\usepackage{amssymb}
\usepackage{amsfonts}
\usepackage{amsmath}

\begin{document}
\renewcommand{\thefigure}{\arabic{figure}}

\title{Spin Drag and Spin-Charge Separation in Cold Fermi Gases}
\date{\today}

\author{Marco Polini}
\email{m.polini@sns.it}
\affiliation{NEST-CNR-INFM and Scuola Normale Superiore, I-56126 Pisa, Italy}
\author{Giovanni Vignale}
\affiliation{Department of Physics and Astronomy, University of
Missouri, Columbia, Missouri 65211, USA}

\begin{abstract}
Low-energy spin and charge excitations of one-dimensional interacting fermions are completely decoupled and propagate with different velocities. These modes however can decay due to several possible mechanisms.   In this paper we expose a new facet of spin-charge separation:  not only the speeds but also the damping rates of spin and charge excitations are different.  While the propagation of long-wavelength charge excitations is essentially ballistic,  spin propagation is intrinsically damped  and diffusive.    We suggest that cold Fermi gases trapped inside a tight atomic waveguide offer the opportunity to measure the spin-drag relaxation rate that controls the broadening of a spin packet.
\end{abstract}
\maketitle

Quantum many-body systems of one-dimensional ($1D$) interacting fermions have attracted an enormous interest for more than fifty years~\cite{giamarchi_book}. Contrary to what happens in dimensionality $D=2$ or $D=3$, these systems cannot be described by the Landau theory of normal Fermi liquids~\cite{Giuliani_and_Vignale}. The appropriate paradigm for $1D$ interacting fermions is instead provided by the ``Luttinger liquid" concept introduced by Haldane in the early 1980s~\cite{haldane}.  The distinctive feature of the Luttinger liquid is that its low-energy excitations are collective oscillations of the charge {\it or} the spin density, as opposed to individual quasiparticles  that carry both charge and spin.  This leads immediately to the phenomenon of {\it spin-charge separation}~\cite{giamarchi_book}, {\it i.e.} the fact that the low-energy spin and charge excitations of $1D$ interacting fermions are completely decoupled and propagate with different velocities.

Recently Recati {\it et al.}~\cite{recati_prl_2003} have proposed to use $1D$ two-component cold Fermi gases~\cite{moritz_2005} to study spin-charge separation. In the case of atoms ``spin"  refers to two internal (hyperfine) atomic states  and ``charge" to the atomic mass density.  Kecke {\it et al.}~\cite{kecke_prl_2005} have pointed out that a giant increase of the separation between charge and spin modes occurs close to the edge of a harmonic potential trap.  Kollath {\it et al.}~\cite{kollath_prl_2005} have performed a time-dependent density-matrix renormalization-group study of spin-charge separation in the $1D$ Hubbard model, giving quantitative estimates for an experimental observation of spin-charge separation in an array of atomic wires.

In this Letter we expose a new aspect of spin-charge separation: namely, spin excitations are intrinsically damped at finite temperature, while charge excitations are not.  The physical reason for  this difference is easy to grasp.  In a traveling spin pulse the up-spin and down-spin components of the current are always equal and oppositely directed, so that the charge density remain constant.  The relative motion of the two components gives rise to a form of friction known, in electronic systems,  as ``{\it spin Coulomb drag}"~\cite{scd_giovanni,scd_flensberg,scd_tosi,weber_nature_2005}.   Of course, no such effect is present in the propagation of charge pulses, which are therefore essentially free of diffusion and damping, at least in the long wavelength limit.  By contrast, a density pulse in a normal Fermi liquid is always expected to decay into electron-hole pairs -- a process known as Landau damping~\cite{Giuliani_and_Vignale}.
In the following we present a quantitative study of this new facet of spin-charge separation, and suggest that the theory could be experimentally tested  in cold Fermi gases.

We consider a two-component Fermi gas with $N$ atoms confined inside a tight atomic waveguide of length $L$ (an ``atomic quantum wire") along the $x$ direction, realized {\it e.g.} using two overlapping standing waves along the $y$ and the $z$ axis as in Ref.~\cite{moritz_2005}. 
The atomic waveguide provides a tight harmonic confinement in the $y-z$ plane characterized by a large trapping frequency $\omega_\perp\simeq 2\pi\times 10~{\rm kHz}$~\cite{footnote_1}. The two species of fermionic atoms are assumed to have the same mass $m$ and different spin  $\sigma$, $\sigma=\uparrow$ or $\downarrow$. The fermions have quadratic dispersion,  $\varepsilon_k=\hbar^2k^2/(2m)$,  and interact  {\it via} a zero-range $s$-wave potential $v(x)=g_{\rm 1D}\delta(x)$~\cite{yang_1967}.

The effective $1D$ coupling constant $g_{\rm 1D}$ is equal to the Fourier transform of the interaction, $v_q$, and can be tuned by using a magnetic field-induced Feshbach resonance between the two different spin states to change the $3D$ scattering length $a_{\rm 3D}$~\cite{moritz_2005}. In the limit $a_{\rm 3D}\ll a_\perp$, where $a_\perp=\sqrt{\hbar^2/(m\omega_\perp)}$, one finds $g_{\rm 1D}=2\hbar^2 a_{\rm 3D}/(ma^2_\perp)$~\cite{olshanii_1998}. In the thermodynamic limit ($N,L\rightarrow \infty, N/L=n$) the properties of the system are determined by the linear density $n$, by the degree of spin polarization $\zeta=(N_\uparrow-N_\downarrow)/N$, and by the effective coupling $g_{\rm 1D}$. The ground-state energy (per atom) $\varepsilon(n, \zeta, g_{\rm 1D})$ can be accurately found by solving a system of Bethe-{\it Ansatz} coupled integral equations (see {\it e.g.} Ref.~\onlinecite{gao_pra_2006}). For future purposes it will be useful to introduce the dimensionless interaction $\gamma_q=m v_q/(\hbar^2 n)$ (for $v_q=g_{\rm 1D}$ this quantity coincides with the usual dimensionless Yang parameter $\gamma$). We also introduce the Fermi wave vector $k_{\rm F}=\pi n/2$, the Fermi velocity $v_{\rm F}=\hbar k_{\rm F}/m$, and the Fermi energy $\varepsilon_{\rm F}=\varepsilon_{k_{\rm F}}$.

The dynamics of density and spin oscillations is controlled by the density-density and spin-spin linear response functions, $\chi_{\rho\rho}(q,\omega)$ and $\chi_{S_zS_z}(q,\omega)$,  which can be conveniently expressed in terms of the symmetric and antisymmetric dynamical local-field factors $G_\pm(q,\omega)$~\cite{Giuliani_and_Vignale},
\begin{equation}\label{eq:lrf}
\left\{
\begin{array}{l}
{\displaystyle \chi_{\rho\rho}(q,\omega)=\frac{\chi_0(q,\omega)}{1-v_q[1-G_+(q,\omega)]\chi_0(q,\omega)}}
\vspace{0.1 cm}\\
{\displaystyle \chi_{S_zS_z}(q,\omega)=\frac{\chi_0(q,\omega)}{1+v_qG_-(q,\omega)\chi_0(q,\omega)}}
\end{array}
\right.\,.
\end{equation}
Here $\chi_0(q,\omega)$ is the noninteracting $1D$ Lindhard response function, which in the long wavelength limit has the form 
\begin{equation}\label{eq:dlimit}
\chi_0(q\to 0,\omega)=\nu(0)\frac{(v_{\rm F}q)^2}{\omega^2-(v_{\rm F} q)^2}\,,
\end{equation}
$\nu(0)=4m/(\pi^2 \hbar^2 n)$ being the $1D$ density of states. The long-wavelength limits of the {\it static} local-field factors  $G_\pm(q,0)$ are fixed by the compressibility and spin-susceptibility sum rules, {\it i.e.}
\begin{equation}\label{eq:static}
\left\{
\begin{array}{l}
{\displaystyle G_+(q\rightarrow 0,0)=1-\frac{\pi^2}{4\gamma}\left(\frac{\kappa_0}{\kappa}-1\right)}
\vspace{0.1 cm}\\
{\displaystyle G_-(q\rightarrow 0,0)=-\frac{\pi^2}{4\gamma}\left(\frac{\chi_{\sigma0}}{\chi_{\sigma}}-1\right)}
\end{array}
\right.\,,
\end{equation}
where $\kappa=n^{-2}\{\partial^2 [n\varepsilon(n,0,g_{\rm 1D})]/\partial n^2\}^{-1}$ and $\chi_{\sigma}=n[\partial^2 \varepsilon(n,\zeta,g_{\rm 1D})/\partial \zeta^2|_{\zeta=0}]^{-1}$ are the compressibility and the spin susceptibility of the interacting system, $\kappa_0=\nu(0)/n^2$  and $\chi_{\sigma 0}=\nu(0)$  are the same quantities for the noninteracting system.

At {\it finite frequency}, however,  the real and imaginary parts of $G_-(q,\omega)$ diverge, in the long-wavelength limit,  as $\omega^2/(q^2v_q)$ and $\omega/(q^2v_q)$ respectively~\cite{qian_prl_2003}.  No such divergence exists in the density channel.  As a result,  the small-$q$ behaviors of the  density-density and spin-spin response functions are dramatically different. 
Indeed, using Eqs.~(\ref{eq:lrf})-(\ref{eq:static}) and the known form of the singularity in $G_-$ it is easy to show that the inverses of these functions have the following form:
\begin{eqnarray}\label{eq:nn_fxcm_lim}
\chi^{-1}_{\rho\rho}(q \to 0,\omega)&=&\frac{m\omega^2}{nq^2}-\frac{m}{n}v^2_{\rm F}\frac{\kappa_0}{\kappa}
\end{eqnarray}
and
\begin{eqnarray}\label{eq:szsz_fxcm_lim}
\chi^{-1}_{S_zS_z}(q\to 0,\omega)&=&\frac{m_{\sigma} \omega(\omega+i\tau^{-1}_{\rm sd})}{nq^2}-\frac{m}{n}v^2_{\rm F}\frac{\chi_{\sigma 0}}{\chi_\sigma},
\end{eqnarray}
where, following the nomenclature introduced in Ref.~\onlinecite{qian_prl_2004},  we have introduced the ``spin mass" $m_{\sigma}$ and 
the inverse of the spin-drag relaxation time $\tau^{-1}_{\rm sd}$ -- both functions of $\omega$ and temperature $T$~\cite{footnote_2}.

\begin{figure}
\begin{center}
\includegraphics[width=1.0\linewidth]{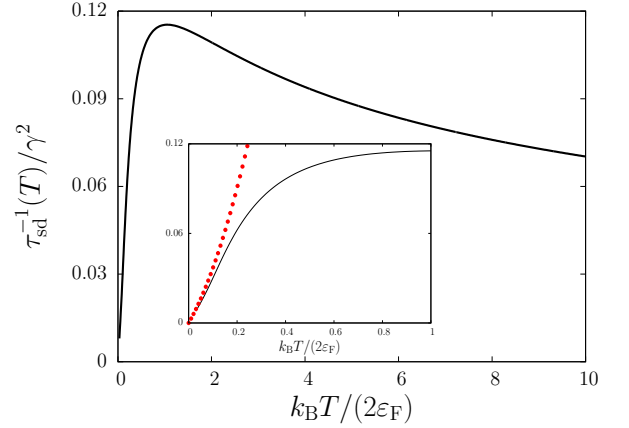}
\caption{(Color online) Spin-drag relaxation rate $\tau^{-1}_{\rm sd}$ (in units of $\varepsilon_{\rm  F}/\hbar$ and divided by $\gamma^2$) as a function of the reduced temperature $k_{\rm B} T/(2\varepsilon_{\rm F})$. In the inset we show a zoom of the low-temperature region $0\leq k_{\rm B}T/(2\varepsilon_{\rm F}) \leq 1$: the filled circles represent the analytical result (\ref{eq:low_t}). \label{fig:one}}
\end{center}
\end{figure}

We now analyze in detail the physical implications of Eqs.~(\ref{eq:nn_fxcm_lim}) and~(\ref{eq:szsz_fxcm_lim}).

According to Eq.~(\ref{eq:nn_fxcm_lim}) the density-density response function has an acoustic pole at $\omega=v_\rho q$ where $v^2_\rho=v^2_{\rm F}\kappa_0/\kappa$. This result is easily recognized to be in agreement with the bosonization result $v_\rho=v_{\rm F} K_\rho\kappa_0/\kappa$~\cite{giamarchi_book}, due to the relation $K_\rho=v_{\rm F}/v_\rho$ that holds for a Galileian invariant system. The speed of sound $v_\rho$ has the following behaviors,  $v_\rho=v_{\rm F}(1+\gamma/\pi^2+...)$ in the weak coupling $\gamma\rightarrow 0$ limit, and  $v_\rho=2v_{\rm F}(1-4\ln{2}/\gamma+...)$ in the strong coupling $\gamma\rightarrow +\infty$ limit~\cite{recati_prl_2003}. Eq.~(\ref{eq:nn_fxcm_lim}) does not incorporate the damping of this acoustic mode associated with the quadratic energy dispersion of the particles: in fact, this damping appears only at higher orders in $q$ and goes to zero as $q^2$~\cite{pustilnik_damping}. 

According to Eq.~(\ref{eq:szsz_fxcm_lim}), the spin-spin response function has an acoustic pole at $\omega=v_{\sigma}q$ where $v^2_{\sigma}=v^2_{\rm F}(m/m_{\sigma})\chi_{\sigma 0}/\chi_{\sigma}$. The bosonization result for the spin velocity is $v_{\sigma}=v_{\rm F}K_\sigma \chi_{\sigma 0}/\chi_{\sigma}$~\cite{giamarchi_book}, with $K_\sigma=1$ due to spin-rotational invariance. The two results for $v_\sigma$ coincide only if the following non-perturbative relation between the spin mass and the spin susceptibility holds:
\begin{equation}\label{eq:identity}
\frac{m_\sigma}{m}=\frac{\chi_\sigma}{\chi_{\sigma 0}}\,.
\end{equation}
In the weak coupling limit $v_\sigma=v_{\rm F}(1-\gamma/\pi^2+...)$. In the strong coupling limit the spin velocity goes to zero as $v_\sigma=2\pi^2 v_{\rm F}/(3\gamma)+...$~\cite{recati_prl_2003}. From these limiting behaviors we find that $m_\sigma/m=1+\gamma/\pi^2+...$ in the weak coupling limit and that the spin mass diverges linearly at strong coupling, $m_\sigma/m=3\gamma/(2\pi^2)+...$.

Let us now examine the spin-drag relaxation time, which is responsible for the damping of the spin mode. In the unpolarized case $N_\uparrow=N_\downarrow$ and within second-order perturbation theory the spin-drag relaxation rate (at zero frequency) is given by the formula~\cite{scd_giovanni}
\begin{eqnarray}\label{eq:nd_order_tau}
\frac{1}{\tau_{\rm sd}(T)}&=&\frac{4 \hbar^2}{n m k_{\rm B} T}
\int_0^{+\infty}\frac{dq}{2\pi}~q^2v^2_q\nonumber\\&\times&
\int_0^{+\infty}\frac{d\omega}{\pi}~\frac{[\Im m \chi_{0}(q,\omega,T)]^2}
{\sinh^2{[\hbar\omega/(2k_{\rm B}T)]}}\,.
\end{eqnarray}
Here $\chi_{0}(q,\omega,T)$ is the $1D$  Lindhard function at finite temperature~\cite{Giuliani_and_Vignale}. 
In Fig.~\ref{fig:one} we plot $\tau^{-1}_{\rm sd}(T)/\gamma^2$ as a function of temperature. We clearly see that the spin-drag relaxation rate goes to zero linearly for $T\rightarrow 0$~\cite{footnote_3}.  Indeed, using Eq.~(\ref{eq:nd_order_tau}) it is possible to show that
\begin{equation}\label{eq:low_t}
\frac{1}{\tau_{\rm sd}(T)}\stackrel {T\to 0}{\rightarrow}\left[\frac{8}{9\pi}\gamma^2_{2k_{\rm F}} \frac{k_{\rm B} T}{2\varepsilon_{\rm F}}
+\frac{8}{3\pi}\gamma^2_0\left(\frac{k_{\rm B} T}{2\varepsilon_{\rm F}}\right)^2\right]\frac{\varepsilon_{\rm F}}{\hbar} \,.
\end{equation}
With increasing temperature the inverse spin-drag relaxation time 
first saturates and then decays to zero rather slowly: $\tau^{-1}_{\rm sd}(T)\stackrel {T\to \infty}{\rightarrow} 16\pi^{-7/2}\gamma^2[k_{\rm B} T/(2 \varepsilon_{\rm F})]^{-1/2}\varepsilon_{\rm F}/\hbar$, for $v_q=g_{\rm 1D}$.

\begin{figure}
\begin{center}
\includegraphics[width=1.0\linewidth]{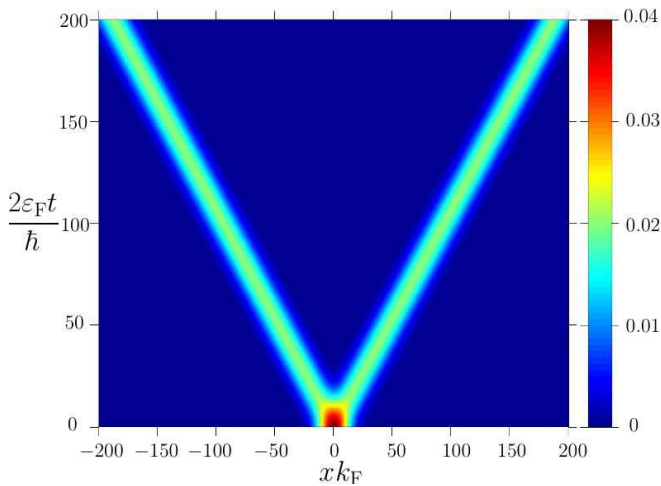}
\caption{(Color online) Space-time evolution of a Gaussian spin packet of initial width $w=10 k^{-1}_{\rm F}$ in the ballistic limit. 
$s(x,t)$ (in units of $k^{-1}_{\rm F}$) is shown as a function of $x$ (in units of $k^{-1}_{\rm F}$) and $t$ [in units of $\hbar/(2\varepsilon_{\rm F})$].
In this example we have chosen $\gamma=0.6$. The left and right components of the packet propagate at essentially the Fermi velocity and their width does not change. \label{fig:two}}
\end{center}
\end{figure}

\begin{figure}
\begin{center}
\includegraphics[width=1.0\linewidth]{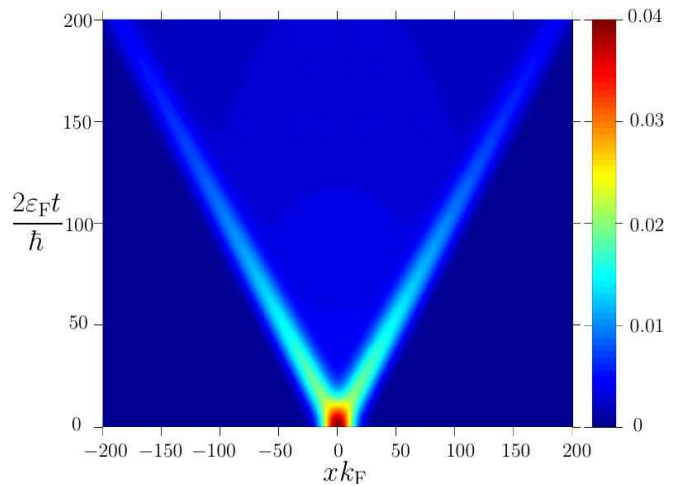}
\caption{(Color online) Space-time evolution of a Gaussian spin packet of initial width $w=10 k^{-1}_{\rm F}$ in the presence of a spin-drag relaxation time $\tau_{\rm sd}\approx 0.04 \hbar/\varepsilon_{\rm F}$ (corresponding to $\gamma=0.6$ and $T=\varepsilon_{\rm F}/k_{\rm B}$).  Notice the spreading and attenuation of the packet as time progresses.\label{fig:three}}
\end{center}
\end{figure}

As discussed in Refs.~\onlinecite{recati_prl_2003} and~\onlinecite{kollath_prl_2005}, localized spin 
and density packets can be created
by short off-resonant and state selective laser light pulses focused to a spot size $w$ such that $R\gg w \gg k^{-1}_{\rm F}$, $R$ being 
the size of the atom cloud and $k^{-1}_{\rm F}$ the average interatomic distance.

The time evolution of these packets is determined by the equations  $\chi^{-1}_{\rho\rho} (q,\omega)n(q,\omega)=0$ and $\chi^{-1}_{S_zS_z} (q,\omega)s(q,\omega)=0$.  Making use of Eqs.~(\ref{eq:nn_fxcm_lim}) and (\ref{eq:szsz_fxcm_lim}) these can be readily converted into two second-order partial differential equations for the density and the spin density:
\begin{equation}\label{eq:differential_equations}
\left\{
\begin{array}{l}
\left(v^{-2}_\rho \partial^2_t -\partial^2_x \right) n(x,t)=0
\vspace{0.2 cm}\\
\left(v^{-2}_\sigma\partial^2_t -\partial^2_x\right) s(x,t)+D^{-1}_\sigma\partial_t s(x,t)=0
\end{array}
\right.\,,
\end{equation}
where we have introduced the spin diffusion constant
$
D_{\sigma}= v^2_{\sigma}\tau_{\rm sd}(T)=n\tau_{\rm sd}(T)/(m_\sigma\chi_{\sigma})
$ (Einstein relation).

The first equation is a 
$1D$ wave equation that depends only on the density velocity $v_\rho$: the solution of this equation with initial conditions $n(x,0)=n_0(x)$ and $\partial_t n(x,t)|_{t=0}=0$ is given by the d'Alembert formula $n(x,t)=[n_0(x+v_\rho t)+n_0(x-v_\rho t)]/2$. The time evolution of a density packet is therefore quite simple: the packet splits into a left-moving and a right-moving component, each one preserving the shape of the initial profile while moving ballistically.

The second equation is a damped 
wave equation that 
depends on two parameters, {\it i.e.} the spin velocity $v_{\sigma}$  and the spin-diffusion constant $D_\sigma$. 
In the undamped $D_{\sigma}\rightarrow +\infty$ limit its solution is given by the d'Alembert expression 
$s(x,t)=[s_0(x+v_{\sigma}t)+s_0(x-v_{\sigma}t)]/2$. In this limit {\it both} density and spin packets move ballistically  (with different velocities). 
This situation is shown in Fig.~\ref{fig:two} for an initial Gaussian spin packet 
$s_0(x)=[(2\pi)^{-1/2} w^{-1}]\exp{[-x^2/(2w^2)]}$ of width $w=10 k^{-1}_{\rm F}$.

For a finite value of the spin diffusion constant the dynamics of the spin packet becomes noticeably different from that of a density packet. 
The solution of the damped 
wave equation satisfying  the initial conditions 
$s(x,0)=s_0(x)$ and $\partial_t s(x,t)|_{t=0}=0$ can be written as
\begin{eqnarray}\label{eq:final_solution}
s(x,t)&=&\exp{\left(-\frac{t}{2\tau_{\rm sd}}\right)} 
\int_{-\infty}^{+\infty}\frac{dq}{2\pi}{\widetilde s}_0(q)\exp{(iqx)}\nonumber\\
&\times& 
\left[\cos{(\omega_q t)}+ \frac{\sin{(\omega_q t)}}{2\omega_q \tau_{\rm sd}}\right], 
\end{eqnarray}
where we have introduced the spin diffusion length  $L_{\sigma}=\sqrt{D_\sigma \tau_{\rm sd}}=D_{\sigma}/v_{\sigma}$, 
$\omega_q=\sqrt{(2qL_{\sigma})^2-1}/(2\tau_{\rm sd})$ (the complex square root is defined here with a positive imaginary part), and    
${\widetilde s}_0(q)=\int_{-\infty}^{+\infty}dx s_0(x)\exp{(-iqx)}$ (Fourier transform of the initial spin density profile). 

\begin{figure}
\begin{center}
\includegraphics[width=1.0\linewidth]{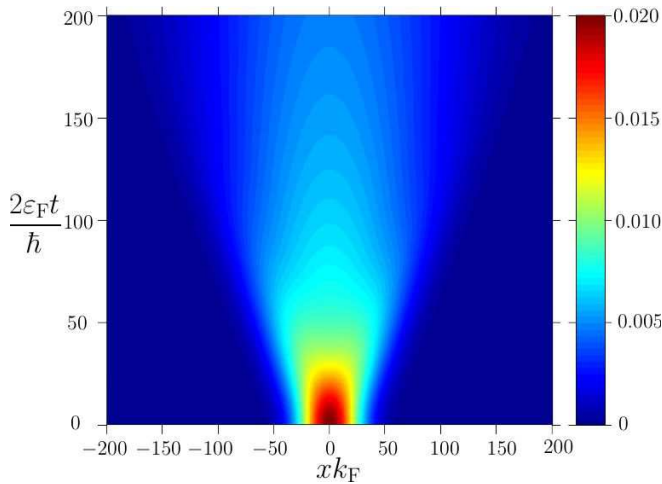}
\caption{(Color online) Space-time evolution of a Gaussian spin packet of initial width $w=20 k^{-1}_{\rm F}$ in the presence of a spin-drag relaxation time 
$\tau_{\rm sd}\approx 0.12 \hbar/\varepsilon_{\rm F}$ (corresponding to $\gamma=1$ and $T=2\varepsilon_{\rm F}/k_{\rm B}$).  The packet spreads out by diffusion and does not split into left- and right-moving components.\label{fig:four}}
\end{center}
\end{figure}

In Fig.~\ref{fig:three} we show the time evolution obtained from Eq.~(\ref{eq:final_solution}) of the same Gaussian spin packet as above 
for $\gamma=0.6$ and $T=\varepsilon_{\rm F}/k_{\rm B}$.  The spreading and attenuation of the packet are quite noticeable.  If the spin-drag relaxation time is sufficiently short ($v_\sigma \tau_{\rm sd}<w$)  it may even prevent the splitting of the initial profile into two peaks:  in that case the evolution of the packet is hardly distinguishable from ordinary spin diffusion in higher dimension (see Fig.~\ref{fig:four}).

In summary, we have shown a new aspect of spin-charge separation in one-dimensional Fermi systems.  Not only spin and charge propagate independently at different speeds -- they are also damped at different rates.  The propagation of charge excitations is essentially ballistic -- any residual damping vanishing in the limit of a smooth density profile.  On the other hand, spin density excitations are intrinsically damped, no matter how smooth the profile, and undergo diffusion, which may completely suppress the ballistic behavior.  We believe that two-component 1D trapped Fermi gases, in which spin pulses can be created and monitored at different times, are ideally suited for an experimental verification of these ideas. 

\acknowledgments 
This work was supported by NSF Grant No. DMR-0313681.


\begin{thebibliography}{77}
\bibitem{footnote_2}
	Although originally introduced for a Fermi liquid, the forms~(\ref{eq:nn_fxcm_lim}) and (\ref{eq:szsz_fxcm_lim}) 
	of the density and spin response functions are generally valid, following from the conservation of total momentum 
	and non-conservation of spin-momentum $P_\uparrow-P_\downarrow$ respectively.
\end{thebibliography}
\end{document}